\RequirePackage{fix-cm}
\documentclass[smallextended]{svjour3}       
\smartqed  
\usepackage{graphicx}
\usepackage{subfigure}
\usepackage[utf8]{inputenc} 
\usepackage[english]{babel} 
\usepackage{amsmath}
\usepackage{amssymb}
\usepackage{txfonts}
\usepackage{graphicx}
\usepackage{placeins}
\usepackage{color}

\font\bbfnt=msbm10

\def\bbR{\mbox{\bbfnt R}}

\newcommand{\fimex}{\hfill$\Box$\end{example}\vspace{0.3cm}}

\newcommand{\bx}{\mbox{\boldmath $x$}}

\begin{document}

\title{Modeling and Analysis of Phase-locked loops: a non reductionist approach }

\author{Jos\'e Roberto C. Piqueira*, Felipe Freitas**  and Luis Antonio Aguirre**}

\institute{\at *Escola Polit\'ecnica da Universidade de S\~ao Paulo \\
              Avenida Prof. Luciano Gualberto, travessa 3, n. 158, 05508-900 \\
              S\~ao Paulo - SP, Brasil \\
              Tel.: 55 11 3091 5647 \\
              \email{jose.piqueira@usp.br}            \\
              ** Universidade Federal de Minas Gerais \\
              Av. Ant\^onio Carlos 6627, 31270-901 \\
              Belo Horizonte - MG, Brasil \\
              \email {aguirre@ufmg.br}
}
\maketitle


\begin{abstract}
Phase-locked loop (PLL), conceived in 1932 by H. Bellescize, has been the basic electronic
component in the development of communication technology from the early analog radio receptors
to modern digital civil and military facilities.
Traditionally, the analysis is conducted by modeling the dynamical behavior of phase and frequency errors,
hence following a phase reduction approach.
One of the main goals of the present work is to describe and investigate the dynamics of
a PLL node by representing it in full state-space, here called non reductionist model,
without the usual design simplifications
i.e., considering different input and output frequencies and not neglecting the higher frequencies components generated in the phase detection process. On the one hand, this approach
complicates the use of analytical tools but on the other hand it permits an efficient numerical
approach that can be used for precise definition of regions in parameters space that show the boundaries between
synchronization and non synchronization regimes, even when noise is considered. Results
show that the PLL node can be simulated in a more realistic way using the state-space model
and that a number of design-relevant aspects can now be investigated numerically.
\end{abstract}



\section{Introduction}

The phase-locked loop (PLL) architecture proposed for frequency demodulation in the 1930's \cite{1}
has been a paradigm for designing a large spectrum of electronic and communication synchronization
strategies, being present in the new generation of wireless devices and dense communication networks \cite{2,3}.
Generally being an anonymous
element, the PLL is a vital element to detect clock signals providing synchronization
for circuits, devices and networks. Despite miniaturization, the original circuit architecture
remained almost unchanged although it improved significantly with digital signal processing.
Such improvement allows accurate and precise operation over a very broad frequency band.
Faster communications and geographical localization are strongly dependent of how clock
signals are distributed and detected. This requires parameter values that take into account
nonlinearities either for the whole network or the interconnected nodes.

Rigorous analytical and practical studies about isolated PLLs \cite{4} and their behavior when
connected forming networks \cite{5,6} revealed that despite all the complexity, PLL digital networks
are efficient and capable of changing our daily life \cite{7}.

Contemporarily, some seminal works appeared \cite{2,3}, presenting mathematical formulations and results
regarding the behavior of the parameters of the entire network,
which served as basic references to the subsequent important developments in communication structures.
Currently, such developments are being used with design tools and
complemented by the analysis of parameter variations \cite{4}.

In this context, the models were developed in a simplified way, considering that the two signals to
be synchronized present almost the same frequency with the terms of phase differences
responsible for the error to be corrected to align the local clock signal with the input coming
from a remote device or network node \cite{8,9}. In such models, the PLL order is defined as
one plus the order of the low-pass filter implemented in the loop \cite{9} and in this work such a
nomenclature is maintained.

Usually, synchronization systems present acceptable performance when second-order PLLs, i.e.\,with
first-order loop filters, are used as network nodes \cite{10}. However there are some situations
where more accurate transient responses are needed, demanding second-order loop filters,
resulting in third-order PLLs \cite{11}.

In such cases, the design must take into account possible instabilities \cite{11}
and, depending on parameter values, Hopf bifurcation and chaotic behaviors can appear \cite{12}.

Unfortunately, phase reductionistic models \cite{13} are not generally effective to choose design
parameters because they do not take into account the high frequency components present at the low pass
filter output even in the second-order case \cite{19} because the filter is considered to be ideal.

Understanding the filter operation in a more realistic way is one of the main motivations
to develop a state-space model for the PLL operation. Other motivations for the non reductionist
model include i)~being able to simulate state variables compatible with measurable magnitudes in real
circuits, hence frequency and phase detection processes can be accurately related to filter parameters;
ii)~the state space model is valid even for large phase errors;
and iii)~it enables the designer to investigate the effect of noise that may appear in different points
of the system. This is a significant advantage over the reductionist model, which is generally noise-free.

This paper aims at investigating dynamical aspects related to synchronization performance
and possible de-synchronization of third-order phase-locked loops (PLL) due to parameters variations.

The model representing the third-order PLL is actually a fourth-order state-space
model where no state variable is the phase, rather the phase must be estimated from the oscillating
signals produced by the PLL components. The state-space model comes much closer to the circuit
implementation of a practical PLL and therefore becomes an important tool for design of such systems.

\begin{figure}[!ht]
\label{fPLLnode}
\centering{
		{\includegraphics[width=.7\textwidth]{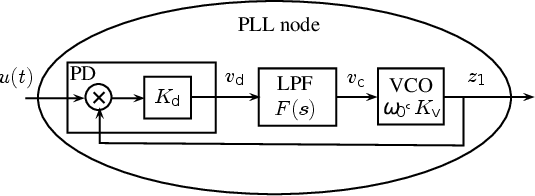}
		\caption {Block diagram of non-reductionist version of a PLL node.
		The $u(t)$ and the node output $z_1$ are multiplied.}}}
\end{figure}

A PLL (Fig.\,\ref{fPLLnode}) is composed of a phase detector (PD) where the input signal,
$u(t)$, and the voltage controlled oscillator (VCO) feedback output, $z_1$, are added and multiplied
by a gain $K_{\rm d}$, resulting the filter input $v_{\rm d}$.

The low-pass filter (LPF) with transfer
function $F(s)$ attenuates high-frequency components, generating the VCO control input $v_{\rm c}$ that adjusts the VCO whole phase, considering that its free-running angular frequency is $\omega_0$, with gain $K_{\rm v}$.

The paper is organized as follows.
In Section~\ref{ssm}, a state space model is studied, with the classical basic
topology of a PLL (Fig.\,\ref{fPLLnode}) used to write the state equations. It is
shown that the local phase PLL model complements the corresponding reductionist version,
reviewed in~\cite{14}, allowing to access the time evolution of the main signals, as illustrated with
an example.
To study how the constitutive parameters of the blocks affect the synchronization performance,
several examples discussed in Sec.\,\ref{sec:perform} confirm the effective contribution of this approach to give a global view of how node parameters affect the synchronization performance. Although the presented approach
is analytically complicated when compared to the traditional model, it permits a simple and accurate
numerical simulations, even considering noise, as exemplified in Section~
\ref{noisy}.
Section~\ref{conc} completes the work with the main conclusion and discussions on how the developed
results can be used according to particular specifications of a design task.


\subsection{Contributions}
The main contribution is the proposition and numerical analysis of a full state-space model instead of the usual
phase reduction model \cite{13}, here called a reductionist model.
The state-space formulation of the PLL node contributes in the design of synchronization systems because
\begin{itemize}
\item nonlinear effects of phase-detection are taken into account;
\item state-space variables are directly related to measurable circuit accessible points;
\item non-ideal filter effects are considered and high frequency disturbances can be evaluated;
\item noise effects can be considered in a friendly way;
\item adequate parameter regions can be estimated, even for a high order filtering process.
\end{itemize}

\section{State space model}
\label{ssm}

Consider the PLL node shown in Fig.\,\ref{fPLLnode}.
The starting point is to write the equation for an oscillator:
\begin{eqnarray}
\label{210322a}
\left\{
\begin{array}{l}
\dot{z}_1 = z_2 \\
\dot{z}_2 = -\omega^2_{\rm inst} z_1,
\end{array} \right.
\end{eqnarray}

\noindent
which has solution $z_1 = A \cos \omega_{\rm inst}t$, where $A$ is a constant that depends on
initial conditions and $\omega_{\rm inst}$ is the instant frequency. A key device in a PLL is the
voltage controlled oscillator (VCO). In order to represent a VCO it suffices to write the instant
frequency as:
\begin{eqnarray}
\label{210322b}
\omega_{\rm inst}(t) = \omega_0 + K_{\rm v} v_{\rm c}(t),
\end{eqnarray}

\noindent
where $\omega_0$ is the central frequency, $K_{\rm v}$ is the VCO gain and $v_{\rm c}(t)$ is the corresponding input
voltage, therefore $z_1 = A \cos [\omega_0 + K_{\rm v} v_{\rm c}(t)]t$, from where it
is seen that the frequency of the VCO given by (\ref{210322a})-(\ref{210322b}) depends on
$v_{\rm c}(t)$.

It is commonplace to normalize the frequency by choosing $\omega_0{=}1$\,rad/s.
Therefore the VCO can be interpreted as implementing a frequency modulated output $z_1(t)$
with carrier $\omega_0$ and modulating signal $v_{\rm c}(t)$.

The voltage $v_{\rm c}(t)$, in turn, is the output of a linear filter, as for instance
\begin{eqnarray}
\label{210322c}
F(s) = \frac{V_{\rm c}(s)}{V_{\rm d}(s)} = \frac{b_1s + b_0}{s^2+a_1s + a_0},
\end{eqnarray}

\noindent
where $V_{\rm c}(s)$ and $V_{\rm d}(s)$ are, respectively, the Laplace transforms of $v_{\rm c}(t)$
and $v_{\rm d}(t)$ which is the output of the {\it phase detector}\, (PD).

$F(s)$ can be realized as
\begin{eqnarray}
\label{210322d}
\left\{
\begin{array}{l}
\dot{x}_1 = x_2 \\
\dot{x}_2 = -a_0 x_1 -a_1 x_2 + v_{\rm d}(t),
\end{array} \right.
\end{eqnarray}

\noindent
with output given by $v_{\rm c}(t) = b_0 x_1+b_1 x_2$.

Clearly, many other filter structures can be
used and since all that is required in the end is to have the corresponding state space realization,
the proposed model can easily accommodate nonlinear filters.

Finally, the phase detector is modeled as the product of the VCO output
and the ``node input'', $u(t)$. Hence
\begin{eqnarray}
\label{210322e}
v_{\rm d}(t) = K_{\rm d} z_1(t) u(t),
\end{eqnarray}

\noindent
where the constant $K_{\rm d}$ is the PD gain.

The final model in state space is:
\begin{eqnarray}
\label{210322f}
\left\{
\begin{array}{l}
\dot{x}_1 = x_2 \\
\dot{x}_2 = -a_0 x_1 -a_1 x_2 + K_{\rm d} z_1 u(t) \\
\dot{z}_1 = z_2 \\
\dot{z}_2 = -[\omega_0 + K_{\rm v}  (b_0 x_1+b_1 x_2)]^2 z_1,
\end{array} \right.
\end{eqnarray}

\noindent
with node input $u(t)$ and node output $z_1(t)$. The second and the forth
equations of (\ref{210322f}) reveal that the node is nonlinear.

As with the physical oscillators, all the state variables are
oscillating signals, that is, none of them is explicitly the output or input phase,
as usually considered. This brings the simulation closer to a
practical situation but at the cost of a more elaborate simulation and
a clear increase in the difficulty of developing any analytical results.

To take $z_1(t)$ as the output is equivalent to use the simple measuring
function $h(\bx)=z_1(t)$, where $\bx \in \bbR^4$ is the state vector.
In Example~\ref{ex1} it will shown that other measuring functions
can be used as, for instance, $h(\bx)=\psi_{\rm o}(t)$, where $\psi_{\rm o}(t)$ is
the output phase.

\begin{example}
\label{ex1}

This example aims at validating model (\ref{210322f}) via numerical integration
using a fourth-order Runge-Kutta with fixed integration interval of $\delta_t=0.01$.
The other parameters used were: $a_0{=}1/3$, $a_1{=}1/2$, $b_0{=}1/3$, $b_1{=}1/12$, that is
\begin{eqnarray}
\label{240322a}
F(s) = \frac{s + 4}{12s^2+6s + 4},
\end{eqnarray}

\noindent
after simple adjustments. The frequency response $F(j\omega)$ is shown in Figure~\ref{PLLBode}.
The gains were taken as $K_{\rm d} {=} K_{\rm v} {=} 0.7$ and $\omega_0{=}1$\,rad/s. The input was
$u(t){=}\sin(1.02t)$ and the initial conditions were taken from a zero-mean
Gaussian distribution with variance $\sigma^2{=}0.01$. It should be noticed that a slight
mismatch between input frequency and the VCO central frequency $\omega_0$ was considered.

\begin{figure}[!ht]
		\centering
		\includegraphics[width=0.7\textwidth]{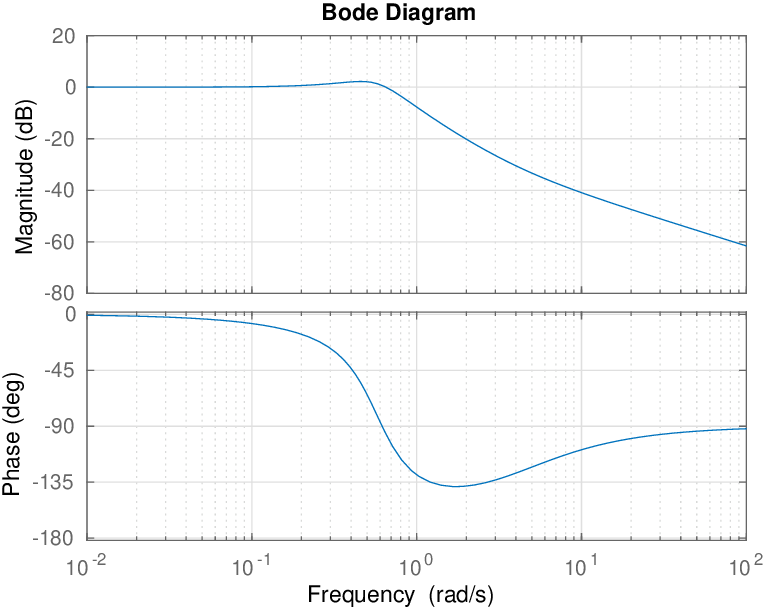}
		\caption{\label{PLLBode}Frequency response of the low pass filter in (\ref{240322a}).}
\end{figure}

The filter input $v_{\rm d}(t)$ and output $v_{\rm c}(t)$ are shown in Figure~\ref{PLLfiltro}. The effect
of the low pass filter is clear from this figure.
It should be noticed that $v_{\rm c}(t)$ oscillates instead of being a dc signal as assumed in the context of
reductionist models~\cite{14}.

To improve filter performance one could reduce the filter
bandwidth or increase the roll-off rate by adding more poles to $F(s)$. The latter alternative is likely
to improve the filter performance but also to deteriorate the PLL performance as the increase of phase-lag
in the loop is known to be deleterious.

\begin{figure}[!ht]
		\centering
		\includegraphics[width=0.7\textwidth]{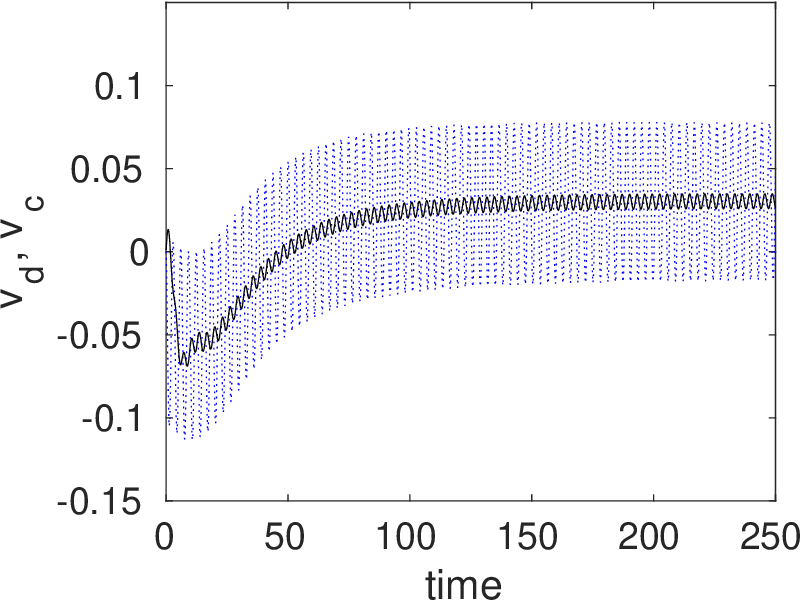}
		\caption{\label{PLLfiltro}The filter input $v_{\rm d}(t)$ is shown in dotted blue lines. This
		has slow- and high-frequency components. The filter output $v_{\rm c}(t)$ is indicated
		in a solid black line and shows that whereas the high-frequency
		component is greatly attenuated the it is not completely eliminated as assumed in the
		reductionist model.}
\end{figure}

The VCO output is shown in Figure~\ref{PLLvco}. The central frequency of this signal is $\omega_0$ and
is roughly half of the high-frequency component of the signal in Figure~\ref{PLLfiltro}.

\begin{figure}[!ht]
		\centering
		\includegraphics[width=0.56\textwidth]{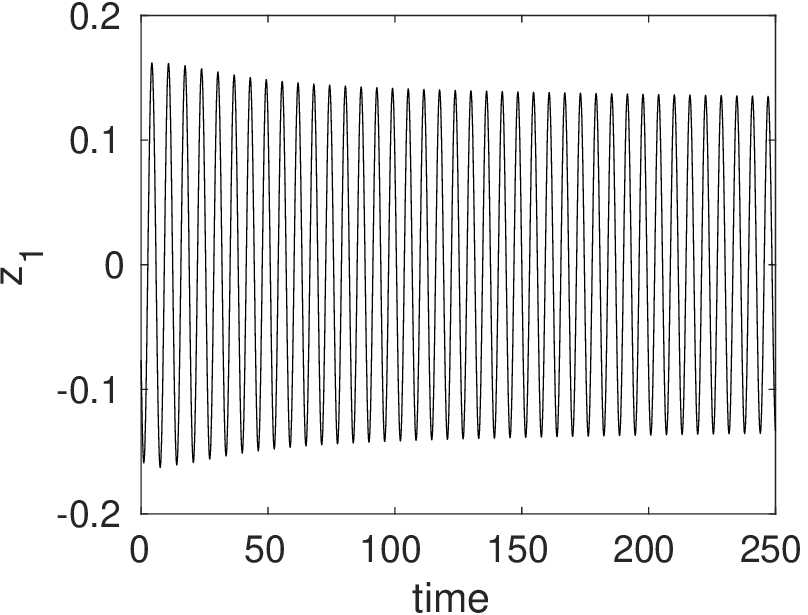}
		\caption{\label{PLLvco}VCO output when the input is solid  black line in Figure~\ref{PLLfiltro}.}
\end{figure}

The question now is how does the phase of the VCO output $z_1(t)$ relate to that of the input signal $u(t)$.

In order to answer this question, Figure~\ref{PLLLisaj} shows a Lissajous plot. If the signals were
identically synchronized, this plot would be a diagonal line. If the signals are phase synchronized,
then the result is an ellipse where the width corresponds to the phase difference

If the phases difference is constant the result is a single ellipse. In Figure~\ref{PLLLisaj}
there are regions in which the line is slightly thicker indicating that the relative phase has a small fluctuation.

\begin{figure}[!ht]
		\centering
		\includegraphics[width=0.56\textwidth]{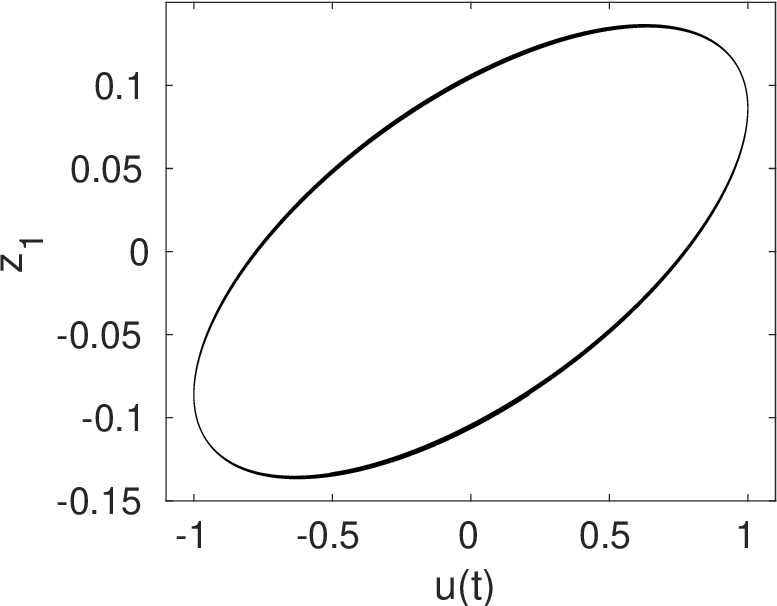}
		\caption{\label{PLLLisaj}This Lissajous plot suggests that the relative phase is nearly
		constant.}
\end{figure}

In many problems it is important to have an estimate of the phase. In the case of the input $u(t){=}\sin(1.02t)$, because
it is sinusoidal, the phase is simply $\psi_{\rm i}(t){=}1.02t$. In the case of the VCO output, a convenient
way of estimating the phase is:

\begin{eqnarray}
\label{240322b}
\psi_{\rm o}(t) = \tan^{-1} \left( \frac{z_2}{z_1} \right) ,
\end{eqnarray}

\noindent
which is in the form of a measuring function $h(\bx)=\psi_{\rm o}(t)$. A relative phase can be defined as
$\theta (t) = e(t) = \psi_{\rm i}(t) -\psi_{\rm o}(t)$. Phase locking happens for constant $\theta (t)$, and the
condition for phase synchronization, which is more relaxed, is $\theta (t)< C$, where $C$ is a constant
usually smaller than $2\pi$.

Figure~\ref{PLLerrofase} shows the first difference of the relative phase, that is,
$e(t)-e(t{-}\delta_t)$, which is an approximation of the derivative of the relative phase.

\begin{figure}[!ht]
		\centering
		\includegraphics[width=0.56\textwidth]{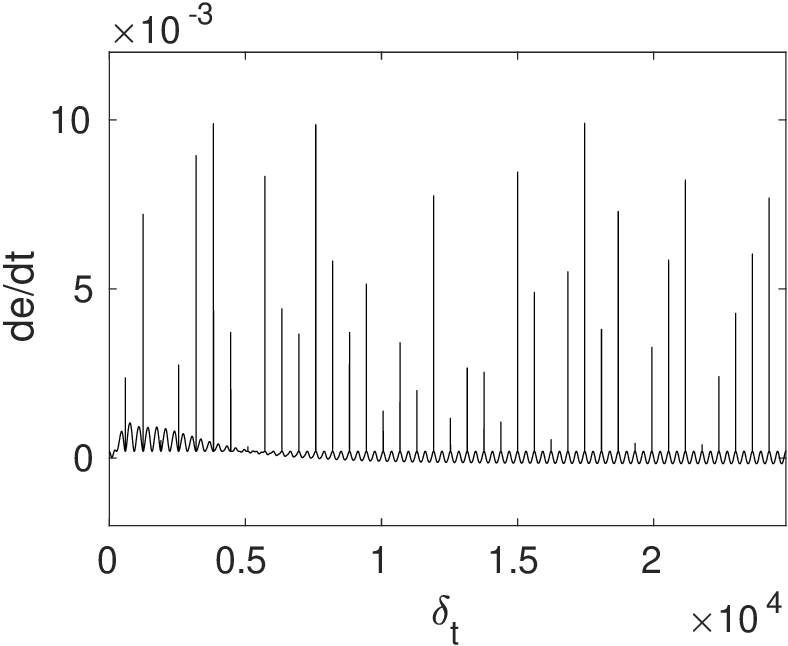}
		\caption{\label{PLLerrofase}First difference $e(t)-e(t{-}\delta_t)$ of the relative phase.}
\end{figure}

As can be seen from Figure~\ref{PLLerrofase} changes in relative phase remain small indicating
phase synchronization. The spikes are probably due to the discontinuity of the {\tt $atan$} function
used to compute the phase. Therefore, despite a small mismatch in the frequency of the input
signal compared to the central frequency of the VCO, the input and output signals of the node
become phase synchronized.
\fimex

Observing the example, non-reductionist strategy to model PLL dynamics provides the possibility
of having the temporal evolution of the periodic signals through the loop, complementing the
reductionistic strategy that gives only corresponding phase and frequency errors
and their trajectory to equilibrium states.

Besides, Figures~\ref{PLLvco},~\ref{PLLLisaj}, and~\ref{PLLerrofase} show signals that have
counterparts in real electronic circuits, differently from the reductionist strategy where
instead of voltage signals they directly provide the phase error.

\section{Performance of the PLL node}
\label{sec:perform}

To investigate the performance of the PLL node as it tries to synchronize
to the node input $u(t){=}\sin \omega_{\rm i} t$, the node phase is computed using equation
(\ref{240322b}), as shown in the next example.

\begin{example}
\label{ex2}

Using the same filter as for Example~\ref{ex1} and the additional parameters: $\omega_{\rm i} {=} 1.001$\,rad/s and
$\omega_0 {=} 1.002$\,rad/s, $\delta_t {=} \pi/300$, $K_{\rm d} {=} K_{\rm v} {=} 0.8$, the VCO
is simulated with (\ref{210322b}) and with an integral term
\begin{eqnarray}
\label{110522a}
\omega_{\rm inst}(t) = \omega_0 + K_{\rm v} v_{\rm c}(t) + K_{\rm i} \int_{0}^{\,t} v_{\rm c}(\tau) {\rm d}\tau,
\end{eqnarray}

\noindent
with $K_{\rm i} {=} 0.5$. It should be noticed that the inclusion of the integral
term in (\ref{110522a}) would not be straightforward in the reductionist model.

The motivation for including the integral term stems from the interpretation
of $v_{\rm c}(t)$ as a proxy for the {\it frequency}\, error and from the aim of achieving null steady-state
error. The main results are shown below in Figures~\ref{Fig110522a} and~\ref{Fig110522b}.

\begin{figure}[!ht]
		\centering
		\begin{tabular}{cc}
		(a) & (b) \\
		\includegraphics[width=0.45\textwidth]{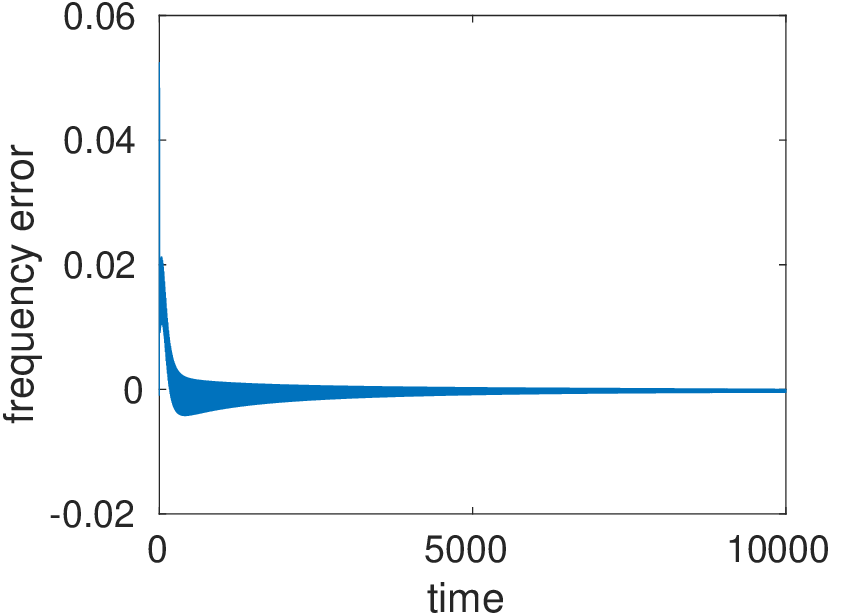} &
		\includegraphics[width=0.45\textwidth]{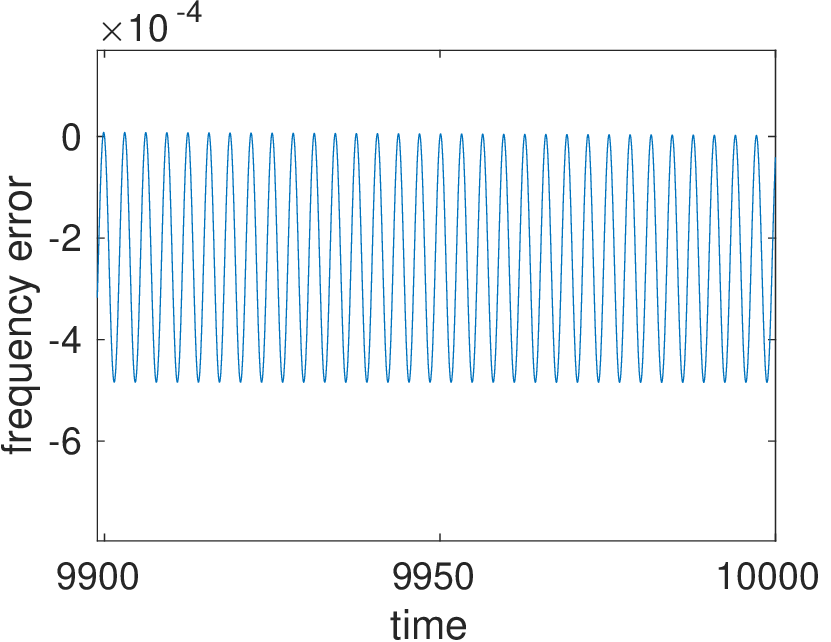}
		\end{tabular}
		\caption{\label{Fig110522a}(a)~Frequency error $\omega_{\rm i}-\omega_{\rm inst}$ over 1,593
		revolutions and (b)~zoom. The VCO is implemented with (\ref{210322b}). Notice that the steady-state error
		is almost always negative.}
		\end{figure}

\begin{figure}[!ht]
		\centering
		\begin{tabular}{cc}
		(a) & (b) \\
		\includegraphics[width=0.45\textwidth]{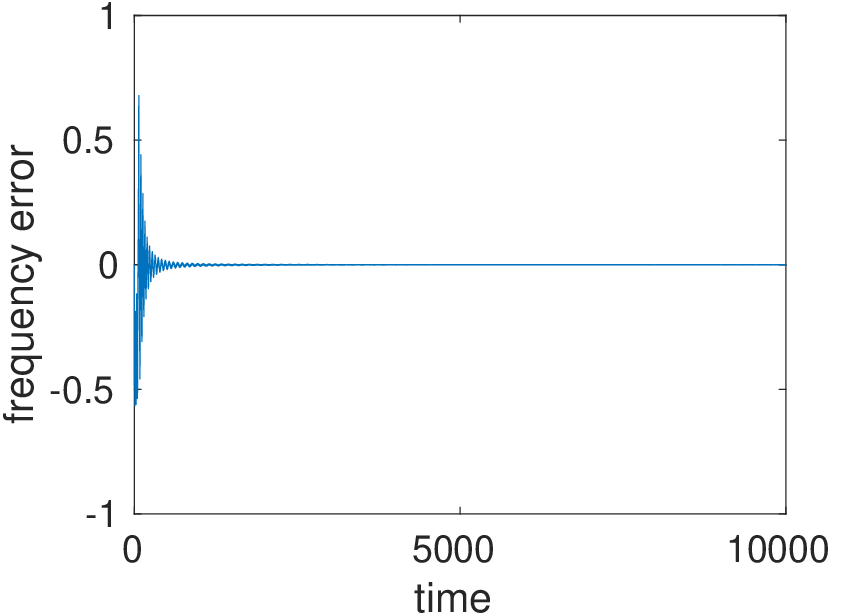} &
		\includegraphics[width=0.45\textwidth]{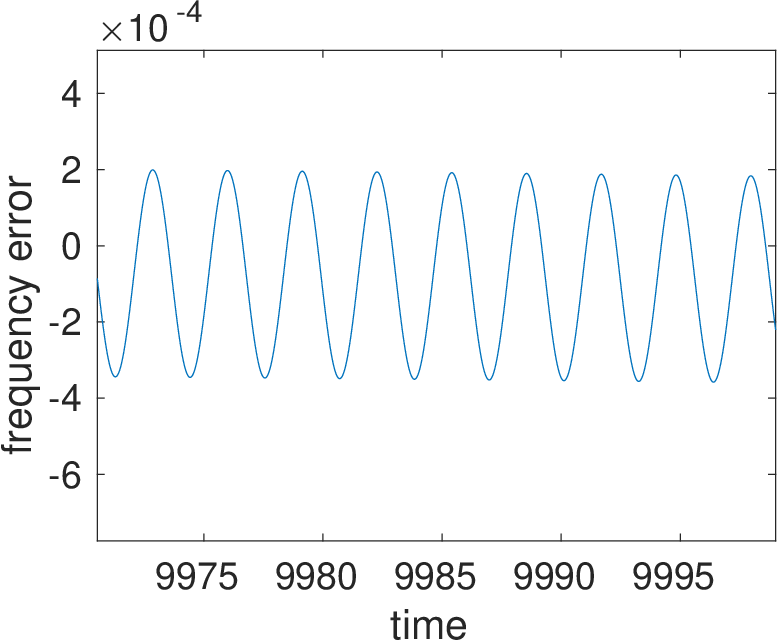}
		\end{tabular}
		\caption{\label{Fig110522b}(a)~Frequency error $\omega_{\rm i}-\omega_{\rm inst}$ over 1,593
		revolutions and (b)~zoom. The VCO is implemented with (\ref{110522a}). Notice that the  error mean
		in steady-state is much closer to zero than for the case in Figure~\ref{Fig110522a}.}
		\end{figure}

%
%

\begin{figure}[!ht]
		\centering
		\begin{tabular}{cc}
		(a) & (b) \\
		\includegraphics[width=0.45\textwidth]{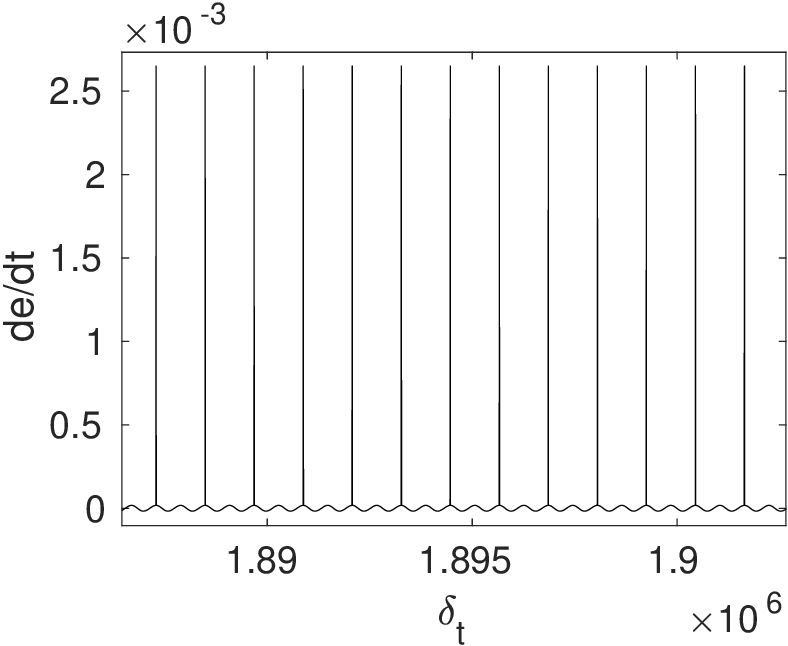} &
		\includegraphics[width=0.45\textwidth]{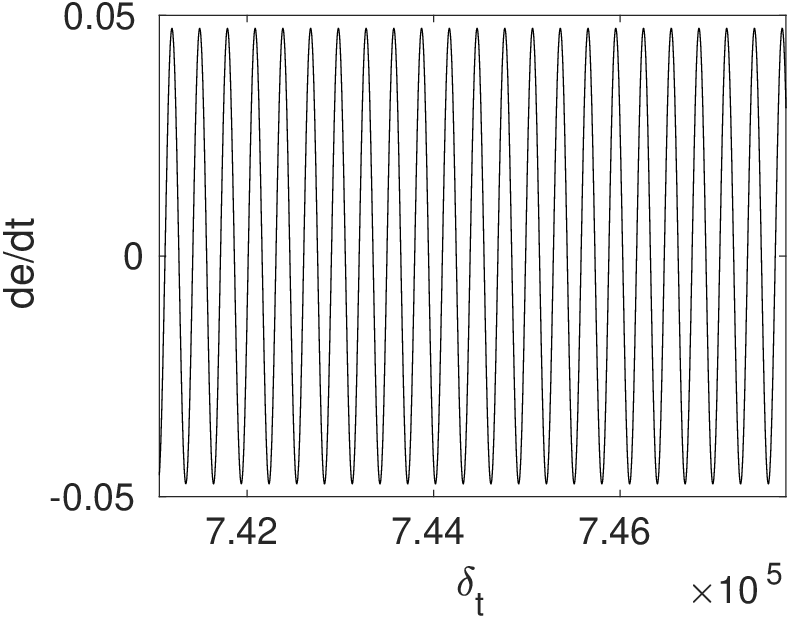}
		\end{tabular}
		\caption{\label{Fig080622}First difference of phase error computed
		(a)~using $e(t){=}\omega_{\rm i}t- \psi_{\rm o}(t)$, see
		(\ref{240322b}) and (b)~using
		$e(t){=}\omega_{\rm i}t- \omega_{\rm inst}t$, see (\ref{210322b}).}
		\end{figure}
\fimex

A common approach to determine if synchronization has or not
been achieved is to define {\it ad hoc}\, thresholds such as \cite{15}[p.\,82]:
\begin{eqnarray}
\label{e310113}
& \left| 1 - \frac{\omega_{\rm i}}{\Omega_{\rm o}} \right | < 2 \times  10^{-3}; \nonumber \\
& | \psi_{\rm o}(t)-\omega_{\rm i}t | < K ,
\end{eqnarray}
where $K$ is a constant often assumed to be equal to $\pi$,
$\Omega_{\rm o}=\langle \omega_{\rm o} \rangle=\langle \dot{\psi}_{\rm o} \rangle$
is the average phase growth rate,
which can be computed as \cite{15}[Eq.\,4.46]
\begin{equation}
\label{c310113}
\Omega_{\rm o} = \lim_{T \rightarrow \infty} \frac{\psi_{\rm o}(T)-\psi_{\rm o}(0)}{T}.
\end{equation}

The second condition in  (\ref{e310113}) was introduced in \cite{16} and is known as the condition for {\it phase entrainment}, which is clearly weaker than condition $\psi_{\rm o}(t)-\omega_{\rm i}t{=}K$. {\it Frequency entrainment}\,  can be tested  using the first condition in  (\ref{e310113}).

In (\ref{e310113}) it is common to take $K=\pi$ because when the phases are {\it not}\, locked, phase
slips of $2\pi$ are verified and hence $\pi$ is sufficiently small to detect such phase slips.

It should be noted
that this is true for most reductionist models where stable and unstable fixed points in the phase relative space
are separated by $\pi$.

In what follows, instead of defining thresholds in an {\it ad hoc}\, way, we compute
\begin{eqnarray}
\label{a230622}
& f = \left| 1 - \frac{\omega_{\rm i}}{\Omega_{\rm o}} \right |;  \nonumber \\
& e = | \psi_{\rm o}(t)-\omega_{\rm i}t |;   \nonumber \\
& m = \langle |\dot{e}| \rangle;  \nonumber \\
& s = {\rm std} (\dot{e} ).
\end{eqnarray}
In (\ref{a230622}), $f$ is a measure of frequency entrainment, $e$ is an estimate of phase error,
where $\psi_{\rm o}(t)$ is computed using (\ref{240322b}), $m$ is the time average of $|\dot{e}|$
and $s$, the standard deviation of~$\dot{e}$.

The reason for using time average
and standard deviation on the derivative of the phase error is that, due to the fact that here the
filtering is {\it not}\, assumed to be ideal, the resulting phase derivative has small oscillations. The following
example illustrates the use of (\ref{a230622}).

\begin{example}
\label{ex3}

Here the filter is the same as for Example~\ref{ex1}, $\omega_0 {=} 1.0$\,rad/s,
$\delta_t {=} \pi/300$ and the VCO is simulated with (\ref{210322b}) and with an integral term
-- see (\ref{110522a}) -- with $K_{\rm i}{=}0.22$. The input frequency was varied in the range
$0.2 \le \omega_{\rm i} \le 1.8$\,rad/s and the gains were within $0.1 \le K_{\rm d} {=} K_{\rm v} \le 3$, yielding 520 different combinations
of $\omega_{\rm i}$ and $K_{\rm d} {=} K_{\rm v}$.

The simulation time for each case was $t_{\rm f}{=}10000$ and the values in (\ref{a230622}) were
computed in the second half of the data to reduce transient effects. Initial conditions were taken
randomly around the origin of the state space. The results are summarized in Figures~\ref{Fig230622bf},
\ref{Fig230622bm} and \ref{Fig230622bs}.

The blue flat floors in Figures~\ref{Fig230622bf} and \ref{Fig230622bm} at the orders of $10^{-3}$ and
$10^{-5}$, respectively,  and correspond to the region in which the PLL node has best performance in terms of
phase-synchronizing with the input. Consequently, as the figure shows, determining PLL
capture amounts to selecting a combination of parameters that is inside de blue region of the diagrams.
This procedure does not require complicated analytical work for the design of higher order synchronization
networks \cite{17}.

The consistency of the synchronization process can be studied observing Figure~\ref{Fig230622bs}
which shows that the variability of the absolute phase error derivative is lower when $\omega_{\rm i} \approx \omega_0$
and gradually increases as  $\omega_{\rm i}$ deviates from the VCO central frequency $\omega_0=1$\,rad/s.
Hence whereas Figures~\ref{Fig230622bf} and \ref{Fig230622bm} can be used to select parameters that result in
synchronization, Figure~\ref{Fig230622bs} goes a step further and shows the subregions, within the blue
plateau, that result in improved robustness of synchronism.

Beyond the boundary of the blue region the performance of the
PLL node degrades abruptly. Roughly
the range of values for which good performance is achieved ``comfortably'' is $0.6 \le \omega_{\rm i} \le 1.5$\,rad/s and
$1 \le K_{\rm d} {=} K_{\rm v} \le 2$.

\begin{figure}[!htb]
		\centering
		\begin{tabular}{cc}
		(a) & (b) \\ \\
		\includegraphics[width=0.45\textwidth]{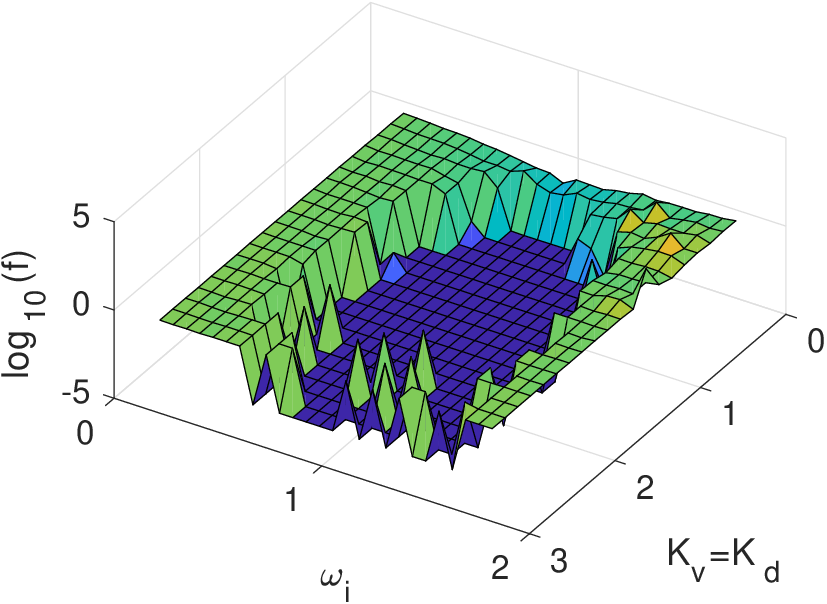} &
		\includegraphics[width=0.45\textwidth]{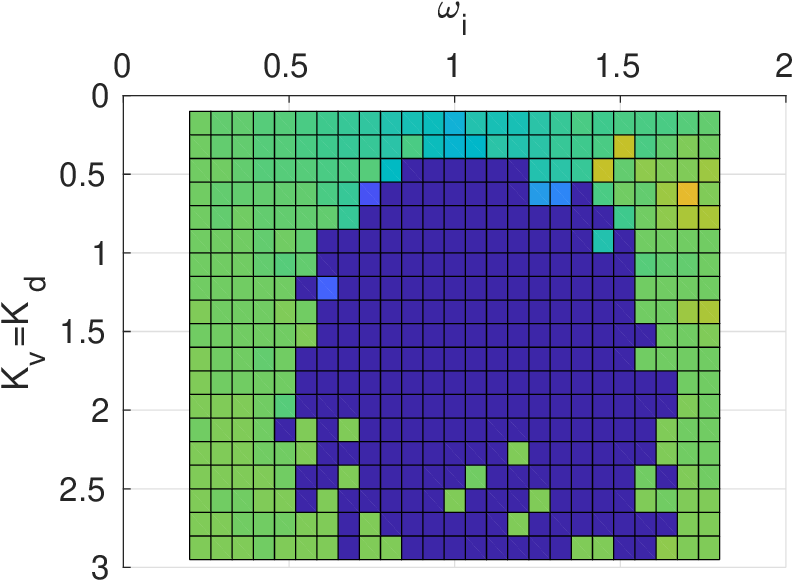}
		\end{tabular}\\
		\caption{\label{Fig230622bf}Logarithm of coefficients in (\ref{a230622}) for Example~\ref{ex3}. The floor in (a)~is at
		the order of $10^{-3}$; (b)~2D top view of (a).}
\end{figure}

\begin{figure}[!ht]
		\centering
		\begin{tabular}{cc}
		(a) & (b) \\ \\
		\includegraphics[width=0.45\textwidth]{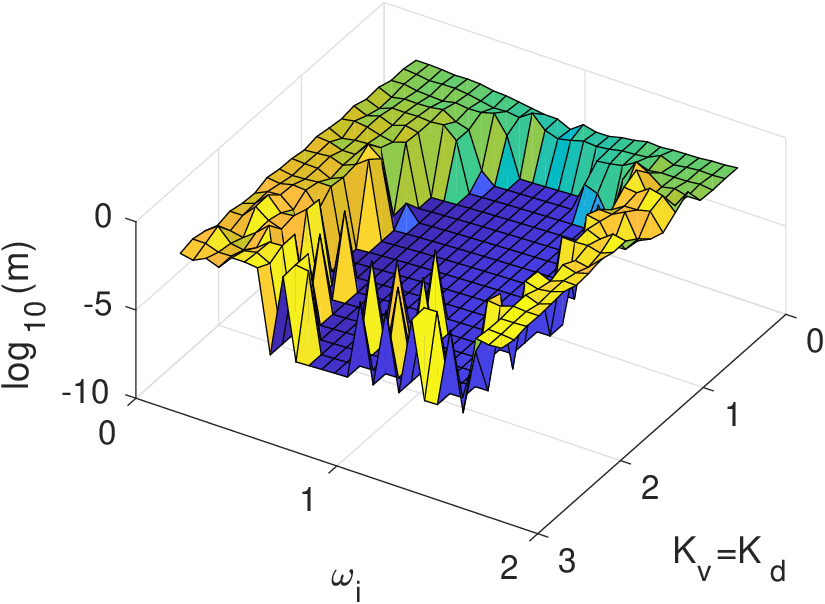} &
		\includegraphics[width=0.45\textwidth]{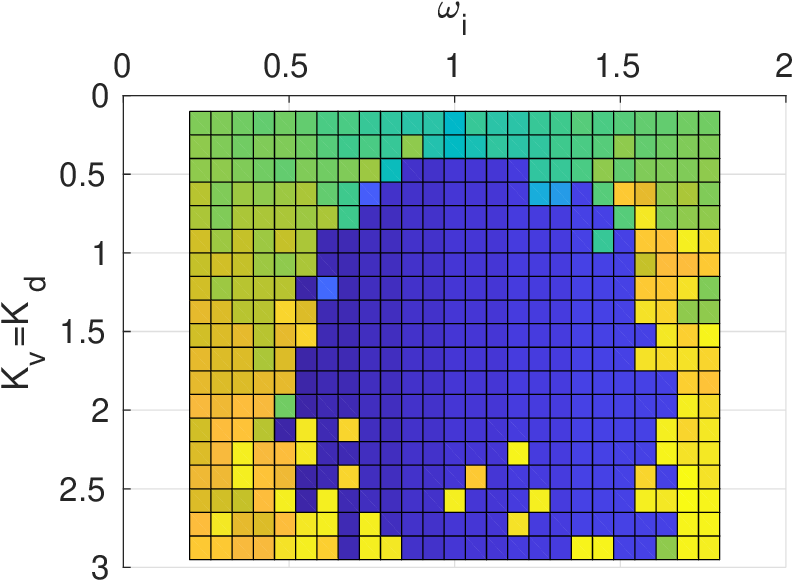}
		\end{tabular}\\
		\caption{\label{Fig230622bm}Logarithm of coefficients in (\ref{a230622}) for Example~\ref{ex3}. The floor in (a)~is at
		the order of $10^{-5}$; (b)~2D top view of (a).}
\end{figure}

\begin{figure}[!ht]
		\centering
		\begin{tabular}{cc}
		(a) & (b) \\ \\
		\includegraphics[width=0.45\textwidth]{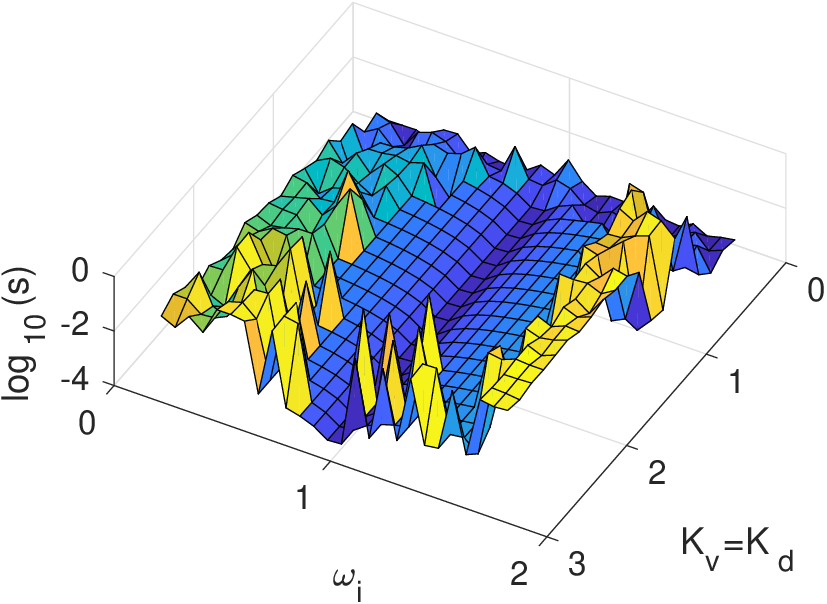} &
		\includegraphics[width=0.45\textwidth]{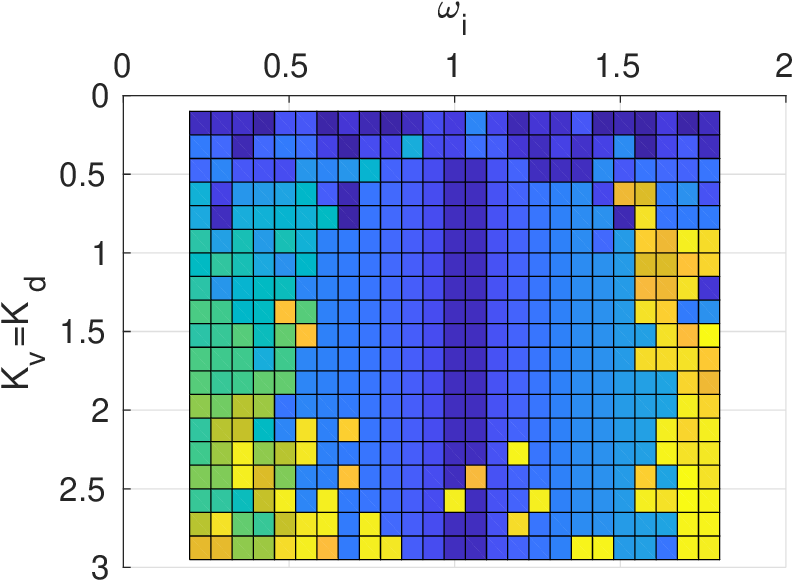}
		\end{tabular}\\
		\caption{\label{Fig230622bs}Logarithm of coefficients in (\ref{a230622}) for Example~\ref{ex3}. The floor in (a)~is at
		the order of $10^{-3.5}$; (b)~2D top view of (a).}
\end{figure}
\fimex
As it can be noticed examining examples 2 and 3, using non reductionist approach to measure PLL performance, despite being analytically complicated, provides
a complete idea about capture and lock-in ranges \cite{8,17} simply observing the permitted synchronization regions of diagrams, allowing to choose gain parameters
according to input frequency variations.

Former works faced the problem of determining capture range for third order PLL by using the reductionist model \cite{11,12,17a} and bifurcation diagrams relating parameters with possible dynamical behaviors.
Although such diagrams present results close to the ones shown in Figures~\ref{Fig230622bf} and \ref{Fig230622bm},
Figures~\ref{Fig230622bf} and \ref{Fig230622bm} are richer than the ones presented in \cite{11,12,17a}
because the transition zones between behaviors are detailed. Additionally, synchronization quality can be evaluated in Figure~\ref{Fig230622bs}.

\section{Noisy Scenarios}
\label{noisy}

An additional feature of the described non reductionist approach is
the capability to determine the capture and lock-in ranges, even in the presence of noise.
In the present study noise is added to the VCO central frequency and
to the input signal and to see how this affects performance and synchronization
boundaries as the gains are varied.

This is illustrated adding noise to some signals in the simulation of the system
described in Example~\ref{ex3}. The noise was taken from a zero-mean Gaussian distribution,
with variance equal to $\sigma^2=0.01$ and added to both: the central frequency $\omega_0$
and the input $u(t)$.

The tuning parameters in this case were $K_{\rm d}=K_{\rm v}=0.8$,
$K_{\rm i}=0$, $\omega_0=1$\,rad/s. The filter parameters are as before
and so was the simulation time $t_{\rm f}{=}10000$.

The initial conditions were taken
randomly around the origin of the state space. The values in (\ref{a230622}) were
computed over the second half of the data.
Figure~\ref{Fignoise290124} summarizes the results when the noise was added to the
VCO central frequency.

As can be seen, the PLL node is quite robust to noise in
$\omega_0$ as about 120 out of 200 runs show performance very close to the noise-free case.
The worse runs indicated in Fig.\,\ref{Fignoise290124}(a) resulted in a frequency mismatch
of about 2.5\%. Over 87\% of the runs had frequency mismatch less than 1\%.

\begin{figure}[!ht]
		\centering
		\begin{tabular}{cc}
		(a) & (b) \\
		\includegraphics[width=0.45\textwidth]{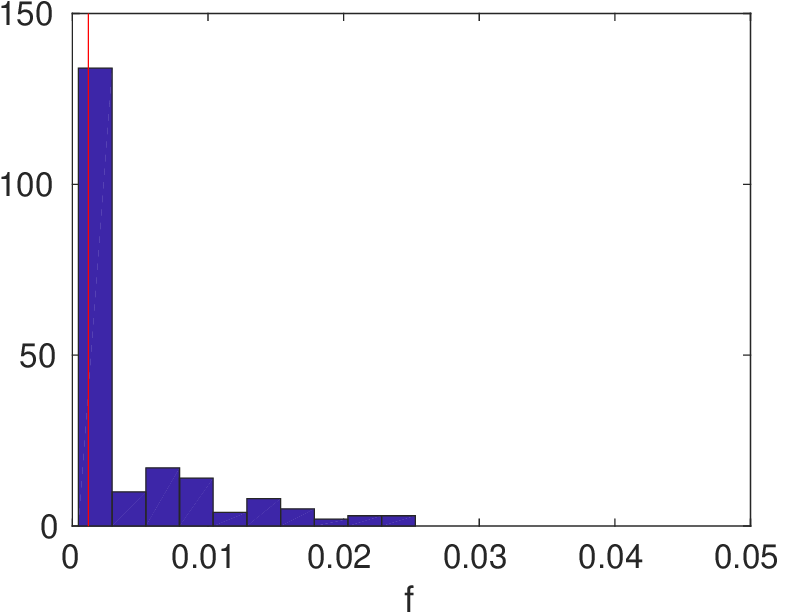} &
		\includegraphics[width=0.45\textwidth]{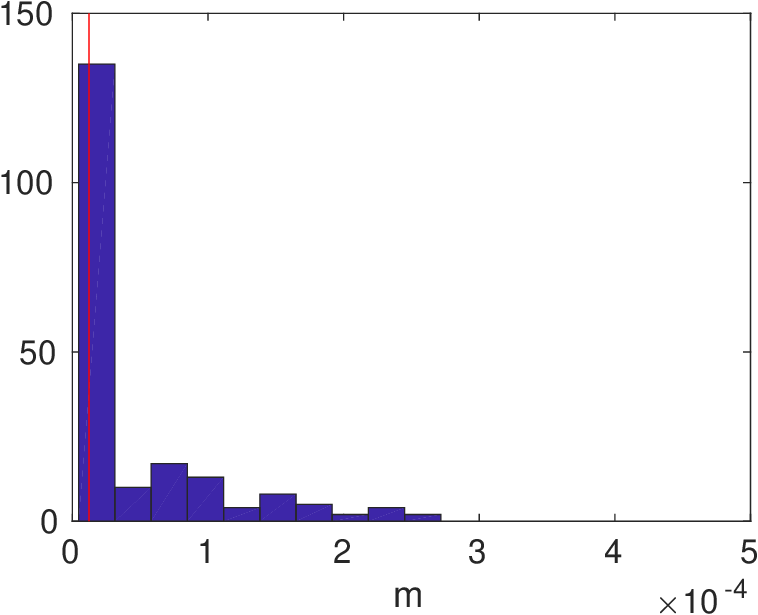}
		\end{tabular}\\
		\begin{tabular}{c}
		(c)\\
		\includegraphics[width=0.45\textwidth]{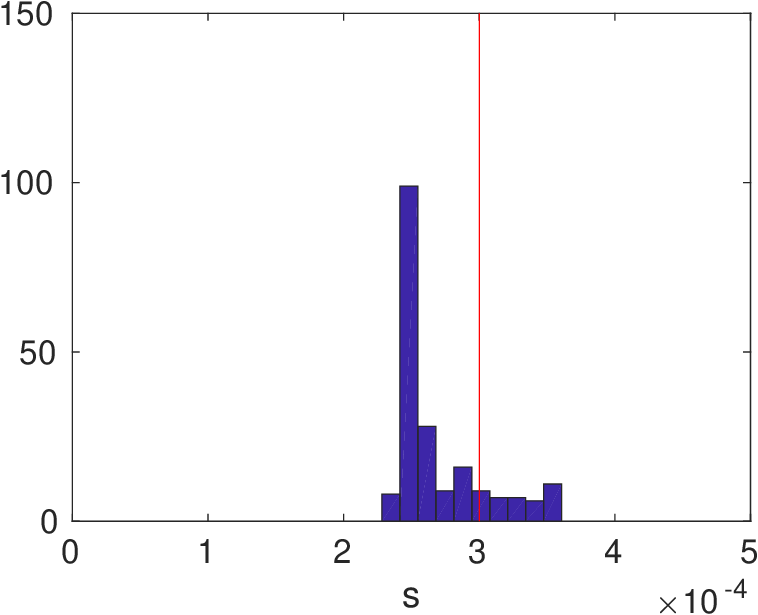}
		\end{tabular}
		\caption{\label{Fignoise290124}Histograms of coefficients in (\ref{a230622}) for 200 Monte
		Carlo runs when noise is added to the parameter $\omega_0$. The red vertical lines correspond
		to the values of (a)~$f$, (b)~$m$ and (c)~$s$ in the noise-free case.}
\end{figure}

Figure~\ref{Figinputnoise290124} summarizes the results when the noise was added to the
input of the PLL node.
As can be seen the robustness in this case is even higher than
for noise in the frequency.
The distributions are very narrow indeed indicating that
the PLL node remains practically unaffected by high-frequency noise in the input.

\begin{figure}[!ht]
		\centering
		\begin{tabular}{cc}
		(a) & (b) \\
		\includegraphics[width=0.45\textwidth]{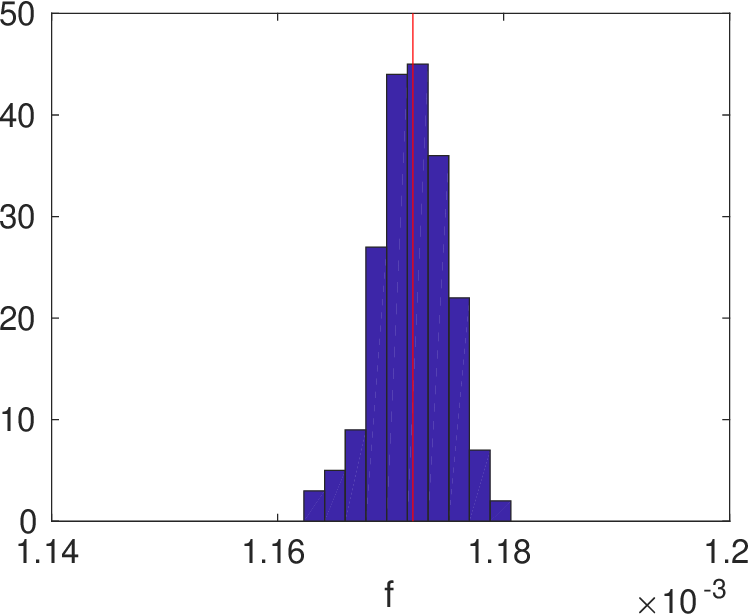} &
		\includegraphics[width=0.45\textwidth]{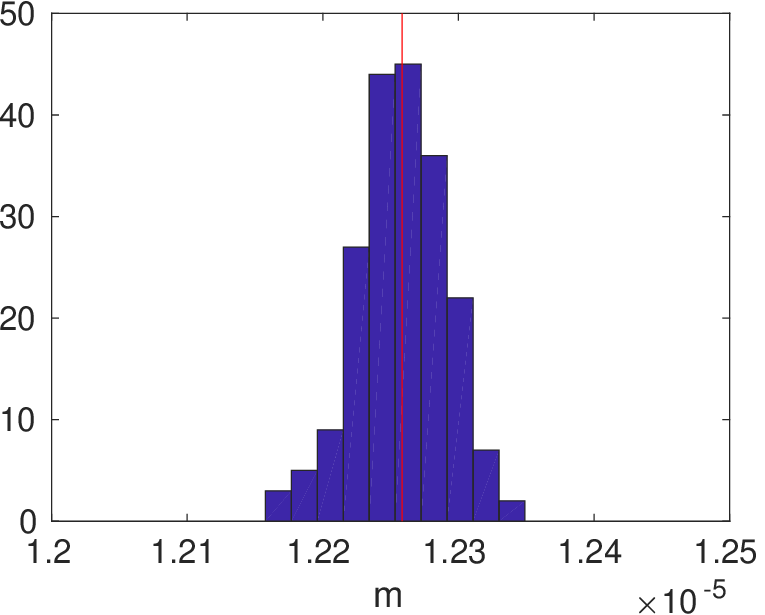}
		\end{tabular}\\
		\begin{tabular}{c}
		(c)\\
		\includegraphics[width=0.45\textwidth]{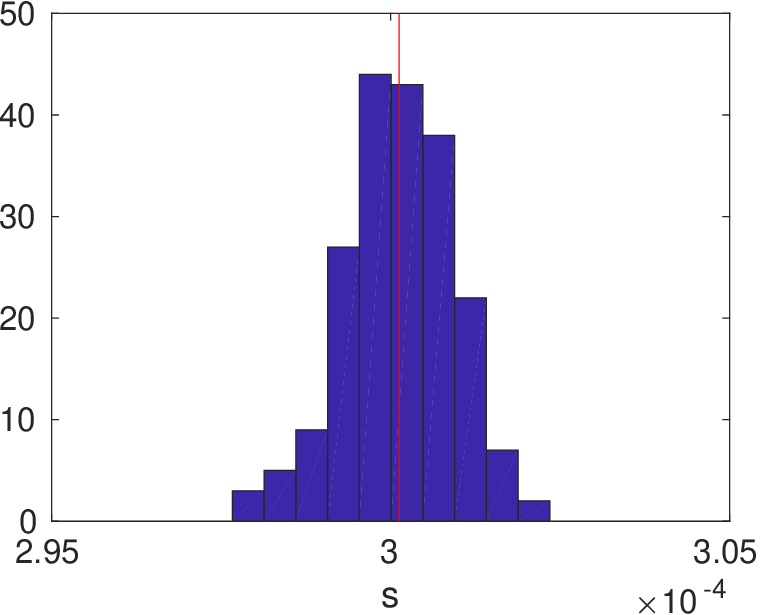}
		\end{tabular}
		\caption{\label{Figinputnoise290124}Histograms of coefficients in (\ref{a230622}) for 200 Monte
		Carlo runs when noise is added to the input signal $u(t)$. The red vertical lines correspond
		to the values of (a)~$f$, (b)~$m$ and (c)~$s$ in
		the noise-free case. Notice that the width of the histograms in this case is much
		narrower than those in Fig.\,\ref{Fignoise290124}.}
\end{figure}

In the sequel, as done in Example~\ref{ex3}, the PLL node was simulated over a grid of 520 combinations of
$K_{\rm v}=K_{\rm c}$ and $\omega_{\rm i}$ values. This time noise was added simultaneously
to both $\omega_0$ and $u(t)$.
The results are summarized in Figures~\ref{Fig300124f}--\ref{Fig300124s} which should be
compared to the noise-free situation, illustrated in Figures~\ref{Fig230622bf},
\ref{Fig230622bm} and \ref{Fig230622bs}.
To facilitate comparison such figures will be
reproduced side by side with the noisy counterpart.

As a general remark concerning Figures~\ref{Fig300124f} and~\ref{Fig300124m}, it is pointed out that the
effect of the noise added simultaneously to $\omega_0$ and $u(t)$ is to reduce the locking region
in parameter space $\{K_{\rm v}{=}K_{\rm c}, \omega_{\rm i}\}$, however whenever synchronization occurs
it is of the same quality as for the noise-free case.
This is seen by noticing that the level of the floor in such figures has the same order of magnitude as
for the noise-free case.

Another interesting remark, which is clearly
seen in the 2D projections is that the noise intensifies the asymmetry of the locking region
especially along the direction of $\omega_{\rm i}$. In other words, the presence of noise limits
the capacity of synchronization for $\omega_{\rm i}<\omega_0$.

Hence during design there is a smaller and more asymmetrical region in parameter space
that will result in synchronization. In practice this means that the choice of parameters in the
noisy case should be done with greater care.


\begin{figure}[!ht]
		\centering
		\begin{tabular}{cc}
		(a) & (b) \\ \\
		\includegraphics[width=0.45\textwidth]{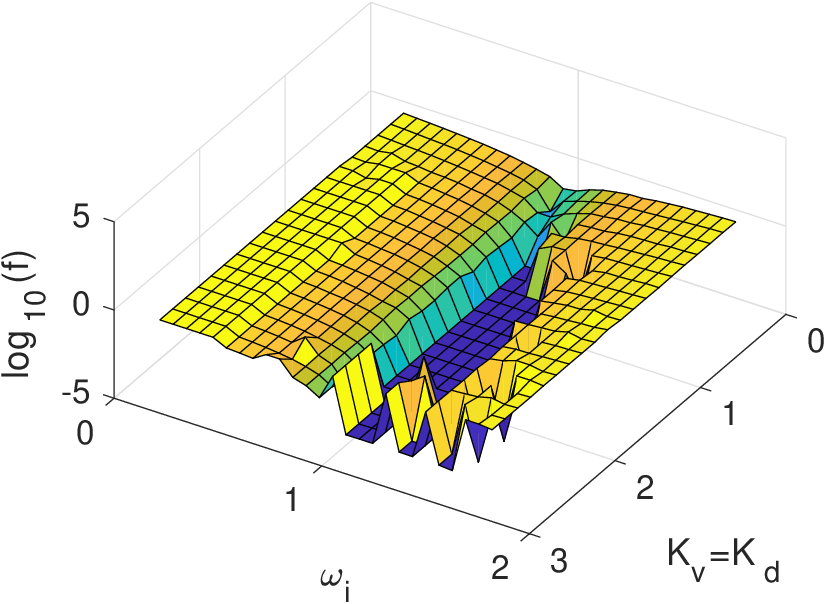} &
		\includegraphics[width=0.45\textwidth]{arnold_f230622b.eps}\\
		(c) & (d) \\ \\
		\includegraphics[width=0.45\textwidth]{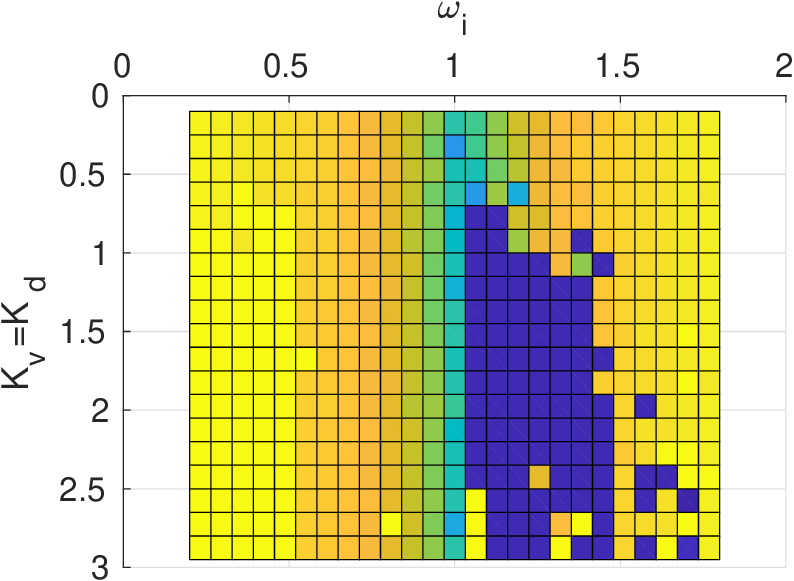} &
		\includegraphics[width=0.45\textwidth]{arnold_f230622b_2D.eps}
		\end{tabular}\\
		\caption{\label{Fig300124f}Logarithm of coefficients in (\ref{a230622}) (a)~and (c)~when noise is simultaneously
		added to $\omega_0$ and $u(t)$; (b)~and (d)~for the noise-free case, see Fig.\,\ref{Fig230622bf}.
		The floor in (a)~and (b)~is at	the order of $10^{-3}$; (c)~and (d)~are the 2D top views of (a)~and
		(b), respectively.}
\end{figure}

\begin{figure}[!ht]
		\centering
		\begin{tabular}{cc}
		(a) & (b) \\ \\
		\includegraphics[width=0.45\textwidth]{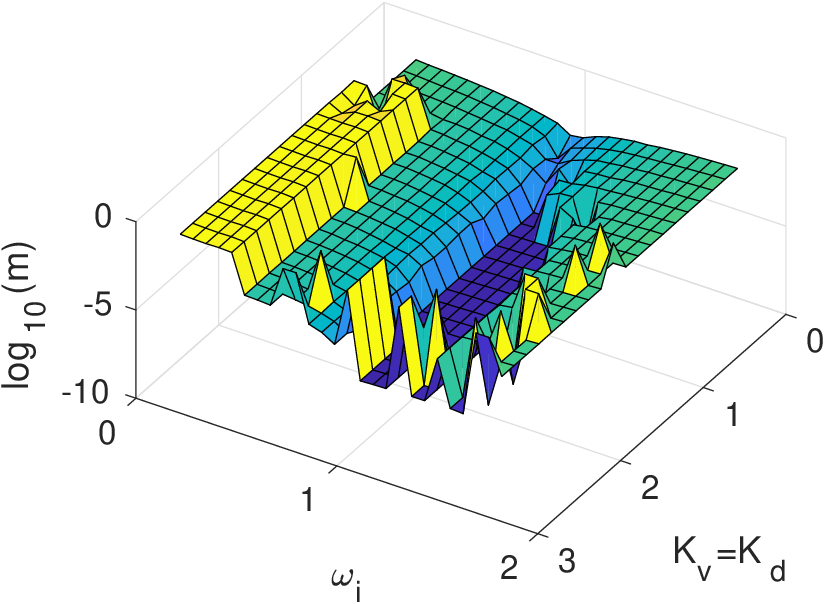} &
		\includegraphics[width=0.45\textwidth]{arnold_m230622b.eps}\\
		(c) & (d) \\ \\
		\includegraphics[width=0.45\textwidth]{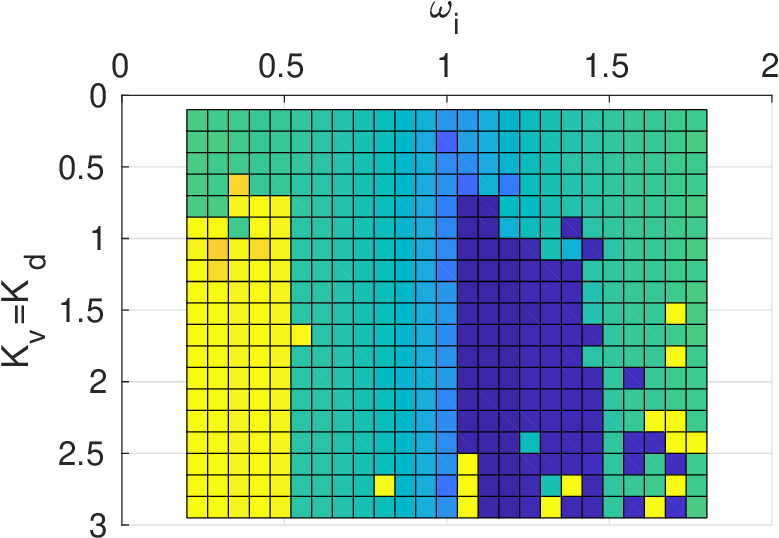} &
		\includegraphics[width=0.45\textwidth]{arnold_m230622b_2D.eps}
		\end{tabular}\\
		\caption{\label{Fig300124m}Logarithm of coefficients in (\ref{a230622}) (a)~and (c)~when noise is simultaneously
		added to $\omega_0$ and $u(t)$; (b)~and (d)~for the noise-free case, see Fig.\,\ref{Fig230622bm}.
		The floor in (a)~and (b)~is at	the order of $10^{-5}$; (c)~and (d)~are the 2D top views of (a)~and
		(b), respectively.}
\end{figure}

A rather curious phenomenon can be observed in Figure~\ref{Fig300124s}, which is a
``stabilizing effect'' of the noise.
Similar situations in other contexts have been reported elsewhere
\cite{18}. It should be noticed that the parameter $s$ -- see (\ref{a230622}) -- quantifies
the dispersion of values of the phase error derivative.
Hence small values of $s$ do not
imply synchronization but rather indicate that whatever is the operating condition the
corresponding variability is small.

\begin{figure}[!ht]
		\centering
		\begin{tabular}{cc}
		(a) & (b) \\ \\
		\includegraphics[width=0.45\textwidth]{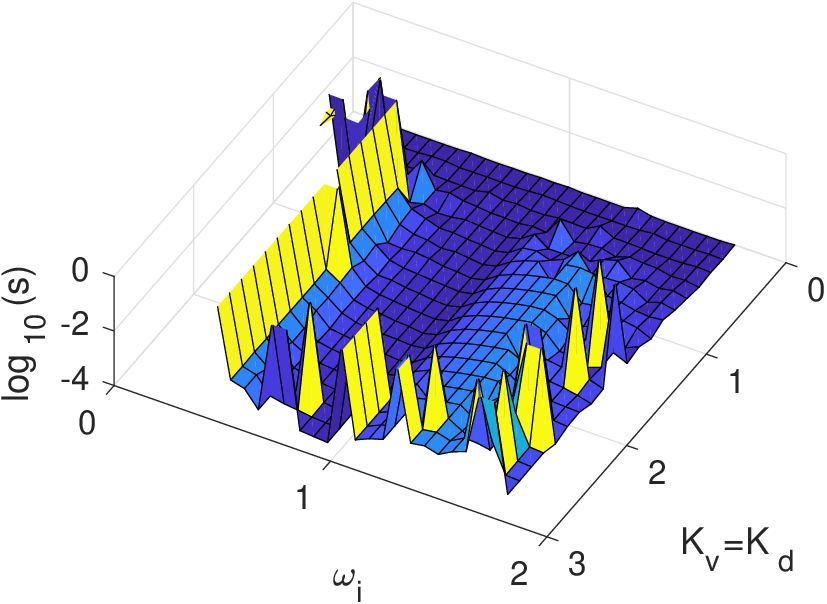} &
		\includegraphics[width=0.45\textwidth]{arnold_s230622b.eps}\\
		(c) & (d) \\ \\
		\includegraphics[width=0.45\textwidth]{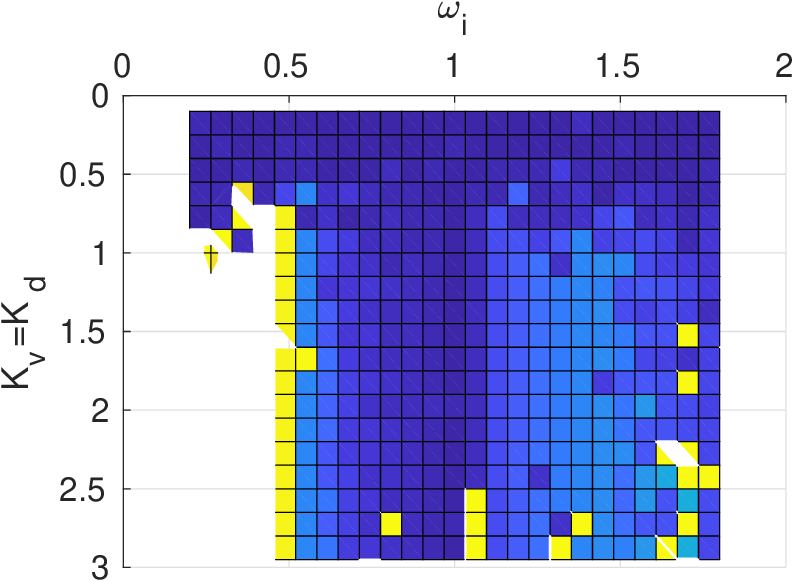} &
		\includegraphics[width=0.45\textwidth]{arnold_s230622b_2D.eps}
		\end{tabular} \\
		\caption{\label{Fig300124s}Logarithm of coefficients in (\ref{a230622}) (a)~and (c)~when noise is simultaneously
		added to $\omega_0$ and $u(t)$; (b)~and (d)~for the noise-free case, see Fig.\,\ref{Fig230622bs}.
		The floor in (a)~and (b)~is at	the order of $10^{-3.5}$; (c)~and (d)~are the 2D top views of (a)~and
		(b), respectively.}
\end{figure}
The analysis of the presented noisy scenarios shows that using the non reductionist model for PLL models is
compatible with Viterbi's seminal work \cite{x3} that has been used for practical design situations \cite{x5}.
A modern view of the problem, as discussed in \cite{x4}, can be considerably improved when phase reduction
approach is replace by the non reductionist one.

\section{Conclusions}
\label{conc}

The main contribution presented is the formulation of the PLL problem in a non reductionist way, differently from the traditional phase reduction approach, by using a dynamic state-space model.

The reductionist model should be used for analysis whereas
the non-reductionist approach should be preferred for more realistic simulations and to make design decisions
which do not rely on the hypotheses underlying the reductionist model that are not verified in practice.
The proposed model, unlike the reductionist one, i)~takes into account the nonlinearity of the sine function,
since the model remains valid for large phase errors; ii)~does not assume the filter is ideal; iii)~enables
simulating and investigation of noisy scenarios; iv)~enables simulating different VCO configurations with
ease, as in (\ref{110522a}).

Therefore with the proposed model friendly numerical analysis can be performed, providing
an accurate view of the capture and lock-in ranges observing regions and boundaries in performance diagrams.

By comparison with studies reported in the literature in seems that abrupt
transitions in the performance diagrams are closely related to bifurcations obtained using
reductionist models that include second-order harmonics.

Consequently, according to the input signal characteristics, bounds for gain parameters adjustments can be easily visualized, even considering noise effects either in the VCO central frequency or in the input signal.

Measures of synchronization performance were defined (Sec.\,\ref{sec:perform}),
allowing to provide criteria to choose gain PLL parameters and, consequently, to set the capture
and lock-in ranges, as shown by examples, operationally important when a clock distribution
network with PLL nodes must be built.

Extensive simulations that consider noise in both the VCO central frequency and in the input signal
show that the state-space model is consistent with some basic features reported in the literature,
such as overall performance when second-order harmonics are not assumed absent \cite{19},
and the pass-band filtering properties of PLLs \cite{4}.

In view of the appealing features of the proposed state-space model, the next step is to
investigate its use in the analysis and design of PLL networks that are typical in
GPS-positioning applications \cite{20}.

\clearpage

\section* {Declaration of competing interest}

The authors declare that they have no known competing financial interests or personal relationships that could have appeared to
influence the work reported in this paper

\section*{Acknowledgments}

JRCP is supported by the Brazilian Research Council (CNPq, grant number: 304707/2023-6) and S\~ao Paulo State Research Foundation (FAPESP, grant number: 2022/00770-0).
LAA is supported by the Brazilian Research Council (CNPq, grant number: 303412/2019-4).

\end{document}